\documentclass[twocolumn,showpacs,preprintnumbers,amsmath,amssymb,prl]{revtex4-1}
\usepackage{graphicx}
\usepackage{amsmath, amsthm, amssymb}
\usepackage{latexsym}
\usepackage{amssymb}
\usepackage{bm}
\usepackage{dcolumn}
\usepackage{epstopdf}
\usepackage{color}

\begin{document}

\title{Microscopic coexistence of a two-component incommensurate spin density wave with superconductivity in underdoped NaFe$_{0.983}$Co$_{0.017}$As}

\author{Sangwon Oh$^{1}$, A. M. Mounce$^{1}$, Jeongseop A. Lee$^{1}$, W. P. Halperin$^{1}$, C. L. Zhang$^{2,3}$, S. Carr$^{2,3}$, Pengcheng Dai$^{2,3}$, A. P. Reyes$^{4}$, P. L. Kuhns$^{4}$}

\affiliation{$^1$Department of Physics and Astronomy, Northwestern University, Evanston, Illinois 60208, USA \\
$^2$Department of Physics and Astronomy, Rice University, Houston, Texas 77005, USA\\
$^3$Department of Physics and Astronomy, The University of Tennessee, Knoxville, Tennessee 37996, USA\\
$^4$National High Magnetic Field Laboratory, Tallahassee, Florida 32310, USAa}
\date{Version \today}

\begin{abstract}
We have performed $^{75}$As and $^{23}$Na nuclear magnetic resonance (NMR) measurements on a single crystal of NaFe$_{0.9835}$Co$_{0.0165}$As and found microscopic coexistence of superconductivity with a two-component  spin density wave (SDW). Using $^{23}$Na NMR we measured the spatial distribution of local magnetic fields.  The  SDW was found to be incommensurate with a major component having magnetic moment ($\sim0.2\,\mu_B$/Fe) and a smaller component with  magnetic moment ($\sim0.02\,\mu_B$/Fe).  Spin lattice relaxation experiments reveal that this coexistence occurs at a microscopic level.
\end{abstract} 

\pacs{ }

\maketitle

The relationship between antiferromagnetism (AFM) and superconductivity in unconventional superconductors, such as cuprates and heavy fermion superconductors, is an intriguing problem of substantial current interest~\cite{nor13,sca12}. In particular, demonstrating coexistence of these two condensations on a microscopic scale is of special importance given the antithetical nature of magnetism and superconductivity~\cite{sat01,lak02,dem01}. Although it has been shown that  iron-based superconductivity~\cite{kam08} can coexist with a spin density wave (SDW)~\cite{pra11,par09,bae09,luo12,wie11,li12,ma12,cai13,ge13,lap09}, in order to better understand this phenomenon it would be helpful to have a clear determination of the spatial distribution of local fields using a high resolution probe. In this Letter we identify an incommensurate SDW that coexists with superconductivity in underdoped NaFe$_{0.983}$Co$_{0.017}$As (NaCo17) taking advantage of very narrow NMR spectra that provide a faithful visualization of the local field distribution.  We find that the SDW has unusual character appearing with two components, one with an amplitude an order of magnitude larger than the other.

\begin{figure}[!ht]
\centerline{\includegraphics[width=0.5\textwidth]{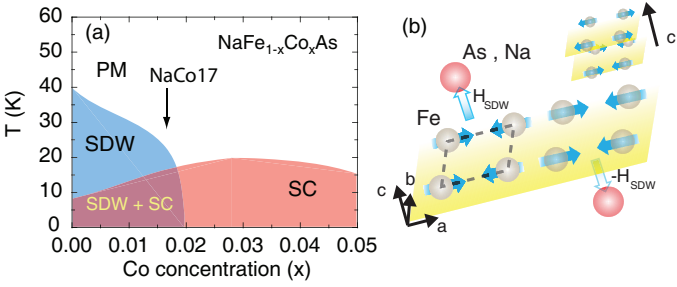}}
\caption{(a) The phase diagram of NaFe$_{1-x}$Co$_x$As as a function of Co concentration~\cite{wan12}. The downward arrow indicates the doping concentration of our Co-underdoped NaCo17 crystal. (b) Alignment of magnetic moments in the SDW state. There is a net alternating magnetic field, $H_{SDW}$, at both As and Na sites along the $c$-axis.}
\label{fig1}
\end{figure}

Theory predicts that coexistence of an SDW and superconductivity is possible where  phases of the  superconducting wave functions on different portions of the Fermi surface with $s_{\pm}$ gap symmetry are different by a factor of $\pi$ for either isotropic or anisotropic superconducting gaps~\cite{fer10b}. It has been argued that an incommensurate SDW is more likely to coexist with superconductivity than for a commensurate SDW~\cite{vor09,vor10b}. In contrast, for $s_{++}$ gap symmetry where the phase of the superconductor  is a constant, coexistence is only possible when the superconducting gap has nodes~\cite{par09b}.  
Nuclear magnetic resonance (NMR) gives a direct measure of  distributions of local magnetic fields,  utilized previously to study SDW's in pnictides~\cite{bae09,li12,ma12}.  Here we investigate a high quality pnictide single crystal, free of paramagnetic impurities, using two resonances, $^{75}$As and $^{23}$Na. The two nuclei are located on opposite sides of the Fe-layer, Fig.\ref{fig1}(b), providing complementary views of the local fields as they develop in the SDW state. The As nucleus has a strong hyperfine coupling to the conduction electrons, in contrast to the Na nucleus, which is much more weakly coupled by a factor of $\sim20$~\cite{oh13}. Owing to the unusually narrow linewidth of $^{23}$Na NMR in NaCo17, $\sim$4 kHz at $T = 30$ K for $H_0 = 16.36$ T, we have determined the spatial distribution of the incommensurate SDW in the normal and superconducting states shown schematically in the phase diagram in Fig.~\ref{fig1}(a).

The $^{75}$As and $^{23}$Na NMR experiments were performed at Northwestern University and the National High Magnetic Field Laboratory in Tallahassee, Florida. The range of temperature was from 2.2 K to room temperature with external magnetic fields from $H_0 = 6.4$ to 16.36 T parallel to the $c$-axis or $ab$-plane. For the results reported here the central-transition linewidth of the $^{23}$Na NMR has been stable at 2.5 kHz for two months, taken at room temperature with $H_0 = 6.4$ T parallel to the $c$-axis; our criterion that there is no degradation of the sample. During the same period the linewidth of the $^{75}$As NMR has remained stable at 8 kHz. The $\sim 3 \times 2 \times 0.3$ mm$^{3}$ crystal of NaFe$_{0.983}$Co$_{0.017}$As was grown at the University of Tennessee and found to have $T_c$ of 18 K from magnetization measurement in low field.  Hahn-echo sequences ($\pi/2$ - $\pi/2$) were used to obtain spectra, and spin-lattice relaxation rates, $1/T_1$, at the central transition (-1/2 $\leftrightarrow$1/2) with a $\pi/2$-pulse of $\approx$ 6 $\mu$sec. A frequency sweep method was used to cover the broadened NMR spectra in the SDW state. $T_1$ was obtained with a saturation recovery method for $T_1<2$ sec. For longer $T_1$ the more efficient progressive saturation technique~\cite{mit01a} was used. 

\begin{figure}[!ht]
\centerline{\includegraphics[width=0.5\textwidth]{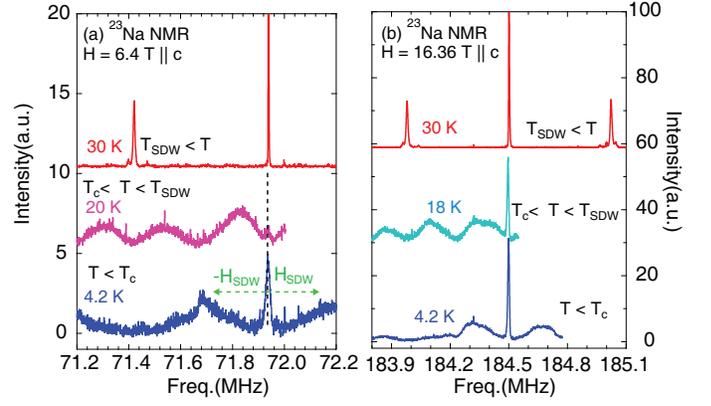}}
\caption{(a) $^{23}$Na NMR spectra from NaCo17 in $H_0 = 6.4$ T parallel to the $c$-axis.  (b) same as (a) but for $H_0 = 16.36$ T.   $T_{SDW}$ and $T_c$ are  27 K and 14 K (12 K) respectively in H = 6.4 T (16.36 T). Two types of SDW were found in the SDW state: a large amplitude SDW (L-SDW) with a splitting in the spectrum $\sim$0.25 MHz (0.02 T), and a small SDW (S-SDW), broadening the spectrum $\sim$ 0.02 MHz (0.0018 T).}
\label{fig2}
\end{figure}

The $^{23}$Na NMR spectra are shown in Fig.~\ref{fig2}. In the paramagnetic state, $T \gtrsim 30$ K, the central and satellite transitions have very narrow linewidths $\approx$ 2.5\,(4) kHz in 6.4\,(16.36) T, even in  high fields, demonstrating  high crystal quality. On cooling, NaCo17 enters into the SDW state at $T_{SDW} \approx 27$ K. Since Na or As atoms are centered symmetrically above or below four Fe atoms, there is a net magnetic field, $H_{SDW}$, at the As or Na sites along the $c$-axis due to in-plane magnetic moments, $m$, at the Fe site~\cite{li09}. Thus a peak in the NMR spectrum at $H_0$ at high temperatures splits into $H_0 - H_{SDW}$ and $H_0 + H_{SDW}$ in the SDW state. $H_{SDW}$ can be expressed as 4 $A_{HF}\, m$ where $A_{HF}$ is the hyperfine coupling constant, $0.023$ T/$\mu_B$ for Na~\cite{oh13,kit11}. This splitting is clearly demonstrated at $T = 18$ and 20 K, Fig.~\ref{fig2}. We could not sweep the complete spectrum owing to available time so the spectra in Fig.~\ref{fig2} (b) above 72 (184.52) MHz at T = 20 (18) K were limited at high frequency and are not shown. From the distribution of the local fields as two broad humps, associated with either the central or the satellite transitions, we can identify  that the SDW is incommensurate. Otherwise each NMR transition would be split into discrete spectra.

Additionally, there is an unexpected peak in the middle of the $^{23}$Na spectrum in the SDW state. We attribute this peak to a small amplitude SDW (S-SDW). The absence of the S-SDW at the satellite position indicates that the electric field gradient (EFG) is  significantly different or disordered at the location of those Na atoms that contribute to the narrow middle peak.  Existence of an SDW below $T_c$ provides evidence for its coexistence with superconductivity, where the transition temperature was determined from coil detuning and spin lattice relaxation. From the known hyperfine field~\cite{oh13} we determined that the moment $m$ for the larger SDW (L-SDW) was  $0.2\,\mu_B$ at $T = 4.2$ K and $H_0 = 16.36$ T.  If we treat the additional contribution to the linewidth of the narrower middle peak as an unresolved splitting we find the moment for the S-SDW to be $0.02\,\mu_B$, under the same conditions.

\begin{figure}[!ht]
\centerline{\includegraphics[width=0.5\textwidth]{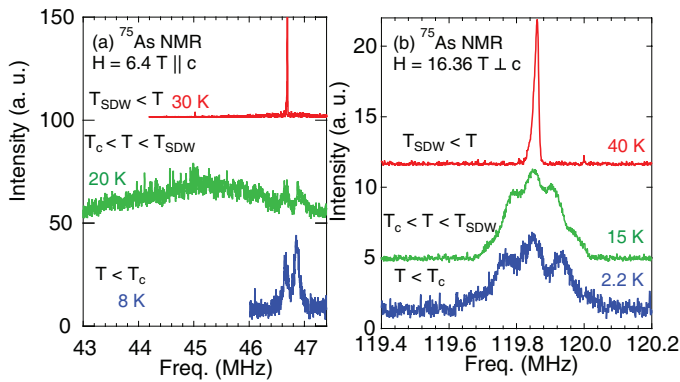}}
\caption{(a)$^{75}$As NMR spectra from NaCo17 in $H_0 = 6.4$ T parallel to the $c$-axis.  The splitting near 46.65 MHz in the SDW and superconducting states indicates that the central peak in the $^{23}$Na NMR is due to a S-SDW not inhomogeneity. (b) $^{75}$As NMR spectra in $H_0 = 16.36$ T $||$ $ab$-plane.}
\label{fig3}
\end{figure}

In Fig.~\ref{fig3}(a) we show $^{75}$As spectra for the paramagnetic phase at $T = 30$ K, in the SDW state $T = 20$ K, and in the superconducting state at $T = 8$ K for $H_0 = 6.4$ T along the $c$-axis.  In the SDW state the wide hump at low frequency for the NMR central transition indicates formation of a large moment SDW. The frequency interval from the unshifted central-transition to the peak of the hump is $\sim$0.2 T (1.5 MHz) at $T = 20$ K, as compared with $^{23}$Na NMR for which the corresponding interval  is $\sim$0.009 T (0.1 MHz). If we allow for the hyperfine field ratio, $A_{HF}^{Na}/A_{HF}^{As} \approx 1/20$~\cite{oh13}, the $^{75}$As and $^{23}$Na NMR consistently depict the same amplitude for the L-SDW at $T = 20$ K.  We also  observe an appearance of two $^{75}$As peaks at the unshifted position with a splitting of 0.034 T (0.25 MHz) that indicates a small  moment incommensurate SDW with ($\sim 0.018\,\mu_B$) at the Fe site, also in-plane, consistent with the broadening of the $^{23}$Na spectra discussed above.  It is interesting that the center of the split spectrum in Fig.~\ref{fig3}(a) does not quite match the peak of the spectrum at $T = 30$ K.  On further cooling, the splitting persists within the superconducting state. We did not perform a frequency sweep to cover the whole range of the L-SDW at T = 8 K due to a drastic decrease in the signal-to-noise ratio in the superconducting state and the rather wide spread in $H_{SDW}$ at the As site.  We note that the small magnetic moment S-SDW is consistent with reports of the amplitude of the SDW observed by neutron scattering on the same material, $\sim0.03\,\mu_B$~\cite{ge13}.

If the external magnetic field, $H_0$, is aligned perpendicular to the $c$-axis, then the net magnetic field at either As or Na sites in the SDW state becomes $H_{net} = \sqrt{H_{0}^2 + H_{SDW}^2}$ where $H_{SDW}$ is parallel to the $c$-axis and its shift in NMR frequency is quadratically suppressed. Thus there is no splitting expected in the SDW state when $H_0$ is strictly parallel to the $ab$-plane.  If there is misalignment of  $H_0$ from the $ab$-plane, $H_{net}$ changes to $\approx H_{0}\pm H_{SDW}\sin\theta$ where $\theta$ is the  misalignment angle.  Then we expect two peaks at $H_0 \pm H_{SDW}\sin\theta$.  Fig.~\ref{fig3}(b) shows $^{75}$As NMR spectra in the normal state ($T = 40$ K), SDW ($T = 15$ K), and superconducting ($T = 2.2$ K) states  where $H_0 = 16.36$ T parallel to the $ab$-plane. In the SDW state and below $T_c$ there are three peaks, two on each side of the unshifted position stemming from the L-SDW and misalignment ($\sim1^{\circ}$), and the central peak appears to be from the S-SDW.  However, in order to distinguish which portion of the spectra corresponds to nuclei affected by superconductivity we have measured spin-lattice relaxation rates, $1/T_1T$.

\begin{figure}[!ht]
\centerline{\includegraphics[width=0.47\textwidth]{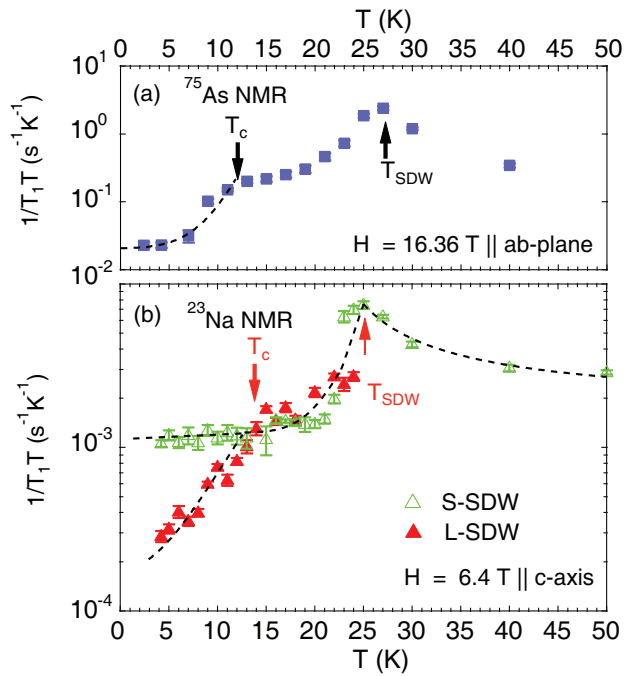}}
\caption{(a) $1/^{75}T_1T$ measured for the S-SDW spectra at $H_0 = 16.36$ T $|| ab$-plane. (b) $1/^{23}T_1T$ at the peak in the spectrum corresponding to each of L-SDW and S-SDW, $H_0 = 6.4$ T $||c$-axis. For both nuclei  a decrease of the rate on cooling through $T_c$  shows  coexistence of the L-SDW with superconductivity on a microscopic scale.  Note that there is minimal change of the transitions with magnetic field in the case of S-SDW from $1/^{23}T_1T$.}
\label{fig4}
\end{figure}

Fig.~\ref{fig4}(a) shows $1/^{75}T_1T$  at the middle peak of the three-peak spectrum in Fig.~\ref{fig3}(b) (S-SDW) with $H_0 = 16.36$ T $\sim$parallel to the $ab$-plane. On cooling $1/^{75}T_1T$ increases due to spin fluctuations~\cite{nin10, oh12,oh13} until $T_{SDW}= 27$ K.  Below this temperature $1/^{75}T_1T$ suddenly drops Arrhenius-like~\cite{sme11}.  Angle resolved photo-emission (ARPES) measurements on a similar NaCo17 sample~\cite{ge13} have shown that there is a remnant density of states at the Fermi surface in the SDW state and consequently $1/^{75}T_1T$  saturates at low temperatures. On further cooling, below $T_c$, the rate decreases more rapidly due to opening of  superconducting gaps at the Fermi surface~\cite{li12,ma12}. Thus the existence of superconductivity at the spatial location of the S-SDW is clear. Additionally, $1/^{75}T_1T$  at either of the two side peaks corresponding to the L-SDW  was found to be the same as the middle peak within $10\%$ at $T = 2.4$ K and consequently we infer that this coexistence with superconductivity exists at a microscopic level uniformly throughout the sample.

We have also performed $1/^{23}T_1T$ measurements at different spectral positions: the central peak (S-SDW) and the hump corresponding to the L-SDW of the $^{23}$Na NMR spectra in different fields along the $c$-axis. The results are shown in Fig.~\ref{fig4}(b). 

\begin{figure}[!ht]
\centerline{\includegraphics[width=0.5\textwidth]{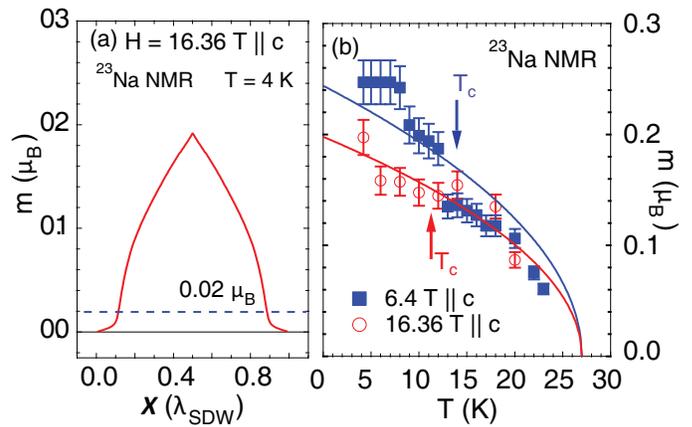}}
\caption{(a) Spatial distribution of magnetic moment, $m$, with period $\lambda_{SDW}$ in the superconducting state calculated from $^{23}$Na NMR~\cite{kit11}. The behavior of the S-SDW  is separated from the L-SDW  by the dashed blue line. (b) The temperature dependence of the most probable magnetic moment for the L-SDW from the $^{23}$Na NMR spectrum is shown for two magnetic fields.}
\label{fig5}
\end{figure}

The spin-lattice relaxation rate  $1/^{23}T_1T$ behaves similar to $1/^{75}T_1T$  increasing on cooling in the normal state and  dropping abruptly below $T_{SDW}= 27$ K and then saturating in the SDW state. However, at the S-SDW position in the $^{23}$Na spectra below $T_c$, $1/^{23}T_1T$  saturates similar to what was noted for optimally doped NaFe$_{0.975}$Co$_{0.025}$As(NaCo25)~\cite{oh13} in contrast to  $1/^{23}T_1T$ at the L-SDW position which drops below $T_c$ similar to $1/^{75}T_1T$.  We speculate that the saturated value for $1/^{23}T_1T$ at the S-SDW corresponds to  \textbf{q} = (0,0) (intra-band) scattering whereas $1/^{23}T_1T$ at the L-SDW is more susceptible to \textbf{q} = ($\pi$,0) or (0,$\pi$) (inter-band) scattering where \textbf{q} is the momentum transfer, similar to optimally doped NaCo25 at low temperature below $T_c$~\cite{oh13}.

We can calculate~\cite{kit11, cur10} the real space distribution of local fields with periodicity $\lambda_{SDW}$ attributed to an incommensurate SDW with magnetic moment $m = H_{SDW}/4A_{HF}$ at the Fe sites using our $^{23}$Na NMR spectra in $H_0 = 16.36$ T.  Fig.~\ref{fig5}(a) shows this spatial distribution as a function of distance, x, in units of the SDW period, where there are two components to $m$. One is  the dominant L-SDW with a large magnetic moment, $m=0.2\,\mu_B$ and the other is the small amplitude SDW (S-SDW) with moment, $m=0.02\,\mu_B$, evident in the foot of the distribution below the dashed line in Fig.~\ref{fig5}(a). The kink at x = 0.5 is an artifact from setting the maximum range of the $H_{SDW}$.  Allowing for  effects of doping, this $m$  corresponds well with the magnetic moments from NMR measurements in undoped NaFeAs where $m$ was found to be $\approx0.3\,\mu_B$~\cite{kit11}.   We also note that the small moment SDW amplitude is comparable to that found in neutron scattering measurements, $\sim0.03\,\mu_B$~\cite{ge13} and we suggest that this might be the dominate contribution in these experiments.   Temperature and magnetic field dependence of the most probable moment from the $^{23}$Na NMR spectrum is shown in Fig.~\ref{fig5}(b). Below $T_c$, the moment slowly increases on cooling in contrast to the decrease reported from neutron scattering measurements, which in that case corresponds to the order of magnitude smaller moment~\cite{ge13,pra11,luo12}. As the ARPES measurements in Co underdoped NaFeAs indicate that significant energy band modification due to the SDW happens at the energy levels which are far from the Fermi energy, more than -50 meV~\cite{ge13}. Thus the discrepancy between the NMR measurement and the neutron scattering can be explained by the different energy excitation scale between the two probes.

 The most natural explanation for the existence of a small moment S-SDW is the disruption of the L-SDW that takes place in the vicinity of a Co atom substituted on an Fe site.  This will suppress the magnetic moment and affect the local $H_{SDW}$  sensed by $^{23}$Na and $^{75}$As NMR. If this were to correspond to  near neighbor and next near neighbor Na positions relative to Co it would affect $\sim 20\%$ of the spectrum in good agreement with our assessment of the S-SDW fraction of the NMR spectra. Furthermore, the S-SDW should be transverse to the L-SDW to coexist with it.  Measurements with well-known and controlled concentrations of dopant will be necessary to confirm this hypothesis. 

In summary, we have performed $^{23}$Na and $^{75}$As NMR experiments to show microscopic coexistence of an incommensurate spin density wave and superconductivity, compatible with $s_{\pm}$-wave  pairing symmetry for NaFe$_{0.983}$Co$_{0.017}$As. The SDW has two components  differing by an order of magnitude in amplitude the smaller of which might be associated with the Co substituting on an Fe site.

\section{Acknowledgments}
We thank Lucia Bossoni and Eric Bauer for useful discussions.  Research was supported by the U.S. Department of Energy, Office of Basic Energy Sciences, Division of Materials Sciences and Engineering under Awards DE-FG02-05ER46248 (Northwestern University). The single crystal growth at the University of Tennessee and Rice University was supported by U.S. DOE, BES under grant No, DE-FG02-05ER46202\,(P.D.)\\

\end{document}